\documentclass[]{spie}  

\usepackage{amsmath,amsfonts,amssymb}
\usepackage{graphicx}
\usepackage[colorlinks=true, allcolors=blue]{hyperref}
\usepackage{multirow}

\usepackage[]{aas_macros}

\graphicspath{{Figures/}}

\title{The final design of GMagAO-X: high-contrast imaging at first-light of the GMT}

\author[a]{Jared R. Males}
\author[a]{Laird M. Close}
\author[b]{Sebastiaan Y. Haffert}
\author[a]{Victor Gasho}
\author[a]{Adam Fletcher}
\author[d]{Joseph Boales}
\author[d]{Dhruv Lakhani}
\author[a]{Maggie Y. Kautz}
\author[a]{Doug Kelly}
\author[a]{Olivier Durney}
\author[a]{Thomas Salanski}
\author[a]{Peter Gray}
\author[d]{Cody Nelson}
\author[d]{Steven Cornelissen}
\author[d]{Paul Bierden}
\author[a,c,e,f]{Olivier Guyon}

\affil[a]{Steward Observatory, University of Arizona}
\affil[b]{Leiden Observatory, Leiden University, The Netherlands}
\affil[c]{James C. Wyant College of Optical Sciences, University of Arizona, USA}
\affil[d]{Boston Micromachines Corporation, USA}
\affil[e]{Subaru Telescope, National Astronomical Observatory of Japan}
\affil[f]{Astrobiology Center, National Institutes of Natural Sciences, Japan}

\authorinfo{Further author information: (Send correspondence to J.R.M.)\\J.R.M.: E-mail: jrmales@arizona.edu}
\begin{document} 
\maketitle

\begin{abstract}
GMagAO-X will be the first-light high-contrast imager on the 25 m Giant Magellan Telescope.   The driving science case for GMagAO-X is characterization of the atmospheres of nearby rocky exoplanets such as Proxima Centauri b.  The revolutionary increase in spatial resolution and sensitivity provided by GMagAO-X will enable detailed study of such planets for the first time.  Additional science cases include: reflected light characterization of mature giant planets; measurement of young extrasolar giant planet variability; characterization of circumstellar disks at unprecedented spatial resolution; characterization of benchmark stellar atmospheres at high spectral resolution; and mapping of resolved objects such as giant stars and asteroids.  These, and many more, science cases will be enabled by a 21,000 actuator extreme adaptive optics (ExAO) system, an integrated coronagraphic wavefront control system with dedicated deformable mirrors, and a suite of imagers and spectrographs.   Science-driven performance requirements for GMagAO-X include achieving a Strehl ratio of 70\% at 800 nm on 8th mag and brighter stars, and exoplanet characterization at planet:star flux-ratios of 1e-7 at 4 lambda/D (26 mas at 800 nm) separation.  GMagAO-X has been added to the GMT project baseline plan and is in the final design phase, aiming to complete FDR in March, 2027.  The instrument is on track to be ready at first-light of the GMT in the mid 2030s. We provide a brief update of the instrument designed to achieve our ambitious performance targets.  
\end{abstract}


\section{INTRODUCTION}
\label{sec:intro}  

The under construction Extremely Large Telescopes (ELTs) offer significant increases in sensitivity and angular resolution.  In particular, the impact on resolved direct imaging of exoplanets will be revolutionary.  The ELTs will enable the search for the signatures of life on exoplanets for the first time, which is an Astro2020 Decadal Survey\cite{2021pdaa.book.....N} Priority Area for astrophysics over the next decade.  In ``Pathways to Habitable Worlds'', Astro2020 found that one of the four key capabilities needed to achieve this is: \textbf{\textit{Ground-based extremely large telescopes equipped with high-resolution spectroscopy, high-performance adaptive optics, and high-contrast imaging.}}\cite{2021pdaa.book.....N}

GMagAO-X is the planned extreme adaptive optics (ExAO)\cite{2018ARA&A..56..315G} coronagraph instrument for the 25 m Giant Magellan Telescope (GMT).  GMagAO-X was conceived to realize the vision of Astro2020 early in the ELT era.  The segment design of the GMT will allow us to deploy mature, well tested DM technology with high actuator density.   GMagAO-X will characterize large numbers of \textit{temperate}, \textit{mature} exoplanets for the first time\cite{2012SPIE.8447E..1XG,2014SPIE.9148E..20M}, including terrestrial \textit{potentially habitable} exoplanets.

\subsection{GMagAO-X Project Status}

Key milestones in the GMagAO-X project to date include 
\begin{itemize}
\item February, 2021: Conceptual design started 
\item September, 2021: Conceptual Design Review passed 
\item December, 2021: Preliminary design started
\item February, 2024: Preliminary Design Review passed 
\item September, 2025: GMagAO-X added to GMT baseline, final design started 
\item September, 2027: Planned Final Design Review
\end{itemize}

The results of the Conceptual Design Review were presented in Males et al.\cite{2022SPIE12185E..4JM}, and the results of Preliminary Design Review were presented in Males et al.\cite{2024SPIE13096E..0YM}. Here we present a brief update on the final design for GMagAO-X.

\section{Final Design Status}

We summarize the key parameters of GMagAO-X in Table \ref{tab:params}.  See Males et al.\cite{2024SPIE13096E..0YM} for a complete list of science requirements from which these parameters are derived.

\begin{table}[h!]
\centering
\footnotesize
\caption{Key Instrument Parameters of GMagAO-X \label{tab:params}}
\begin{tabular}{|l|c|c|c|l|}
\hline
Parameter   &    Requirement    &    Goal  & Stretch Goal   &  Notes \\
\hline
\hline
Wavelength Coverage & 600 -- 1900 nm & 450 -- 1900 nm & 350--1900 nm & \multirow{2}{3.5cm}{GMAGX-SCI-002, GMAGX-SCI-006} \\
&&&&\\
\hline
Spatial Resolution  & 4.9 mas & 3.7 mas & 2.8 mas & GMAGX-SCI-002 \\
\hline
\multirow{3}{*}{Spectral Resolution} & BB: 10\% bands  &        & & \multirow{3}{3.5cm}{GMAGX-SCI-003, GMAGX-SCI-004, GMAGX-SCI-005}\\
                                    &  Low: 20--100             &          & & \\
                                    &  High: 1000--65,000        & 100,000  & & \\
\hline
Guide Star I Magnitude Range & -1.5 -- 13 & -1.5 -- 15  &  & GMAGX-SCI-007 \\
\hline
Field of View  & 3'' x 3'' & 3.5'' x 6'' &  & GMAGX-SCI-010 \\
\hline
\multirow{3}{*}{Coronagraph Contrast}  & \multirow{3}{*}{1e-7 @ 4$\lambda/D$} & 1e-7 @ 2$\lambda/D$ &  1e-8 @ 1$\lambda/D$  & \multirow{3}{3.5cm}{5$\sigma$ planet:star flux. GMAGX-SCI-001, GMAGX-SCI-008}\\
                                       &                     & 1e-8 @ 6$\lambda/D$ &  1e-9 @ 5$\lambda/D$  & \\
                      &&&&\\
\hline

\end{tabular}
\end{table}

\subsection{OptoMechanical Design}

GMagAO-X is intended to occupy a Folded Port (FP) on the Gregorian Instrument Rotator (GIR) of the GMT.   It consists of a two-level optical table, which will be air-floated to provided vibration isolation. The GIR rotates and tips with elevation of the telescope, and as such the FPs are not gravity invariant. To enable the gravity invariance required for air isolation, GMagAO-X is designed to counter rotate. When GMagAO-X is in operation the GIR will be locked so that GMagAO-X will be oriented along the elevation axis of the GMT.

GMagAO-X will also supply a dedicated custom tertiary mirror (M3).  After M3, a relay includes a fast-steering mirror (FSM) and then couples the beam to the gravity invariant main section of the instrument.

\begin{figure} [h!]
\centering
\includegraphics[width=4.5in]{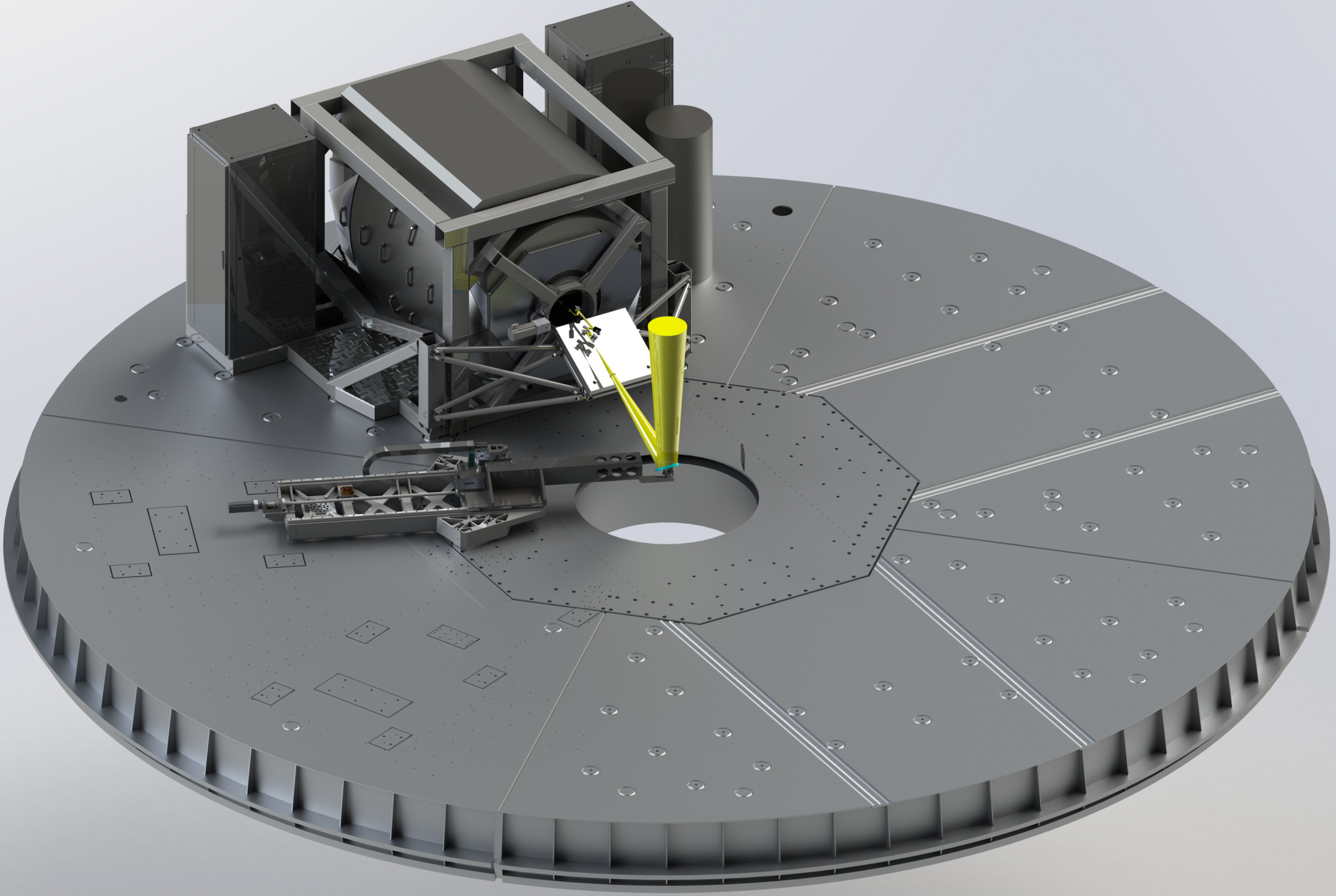}
\caption{\label{fig:mech1} The Mechanical Design of GMagAO-X.}
\end{figure}

After the fore-optics, the beam passes through a pupil alignment periscope, a k-mirror field rotator to keep the pupil alignment fixed, and an atmospheric dispersion compensator (ADC)\cite{giorgetti_spie_2026}.  

A ``woofer'' deformable mirror (DM) is used to provide large-stroke for correcting low-order aberrations at lower speeds.  At PDR a 3228 actuator DM was planned for the ``woofer'', however we are currently reevaluating the trade off between number of actuators, actuator stroke, and temporal response.

After the woofer, the beam is relayed to the 21,000 ``tweeter`` DM which provides the high-speed and high-spatial-frequency control.  A key lesson-learned from on-sky operation of MagAO-X\cite{2024SPIE13096E..0YM} is that stable high-Strehl wavefronts are required for achieving the highest performance from the coronagraph WFS and control (WFS\&C) system. To achieve a projected actuator pitch similar to MagAO-X ($\sim$14 cm), $\sim$3000 illuminated actuators are needed per segment.  GMagAO-X will achieve this with a unique ``Parallel DM'' architecture, consisting of seven 3000 actuator MEMS deformable mirrors.  A hexagonal prism with its 6 segments coated as mirrors splits the GMT aperture such that each segment is imaged onto a 3K MEMS DM.  The central hole passes the center segment to its DM in the back.  The parallel DM has been prototyped and demonstrated, albeit without MEMS installed, in the High Contrast Adaptive Optics phasing Testbed (HCAT) at the University of Arizona\cite{2024JATIS..10d9005K}. After the ``tweeter'', the beam is split between the wavefront sensors (WFS) and the coronagraph.  The WFS system is described below.

The coronagraph beam is relayed to the upper level and to a 3000 actuator non-common path correcting (NCPC) DM, an architecture which has been demonstrated on-sky with MagAO-X\cite{2024SPIE13096E..0YM}. The coronagraph is a Lyot-style architecture which will support a range of designs, including the Phase-Apodized Pupil Lyot Knife-Edge Coronagraph (PAPLKEC)\cite{2020ApJ...888..127P}, and the Phase Induced Amplitude Apodization (PIAA)\cite{2003A&A...404..379G} coronagraph with transmissive complex focal plane masks.  

Focal plane instrumentation such as imagers covering the optical through 1 $\mu$m, and near-IR through H band as well as spectrographs are included. For highest efficiency we are designing an on-board lower-resolution IFU. It may be possible to couple GMagAO-X to the facility G-CLEF\cite{2024SPIE13096E..0ZS} and GMTNIRS spectrographs as a fiber fed integral field unit (IFU).

See Close et al.\cite{close_spie_2026_2} in these proceedings for further details about the optical design of GMagAO-X.

\subsubsection{Deformable Mirror Development}

The required stroke for an AO system is proportional to $(D/r_0)^{5/6}$ \cite{1976JOSA...66..207N}.  MagAO-X has 3.5 $\mu$m of total surface stroke in its ``tweeter'', and 15 $\mu$m of tip/tilt stroke in its woofer.  Moving from a 6.5 m to a 25.4 m telescope means that up to a factor of 3 more stroke is needed (mitigated somewhat by the outer scale).  The FSM in the fore-optics of GMagAO-X alleviates the most demanding tip/tilt stroke requirements. 

\begin{figure} [h!]
\centering
\includegraphics[width=3.5in]{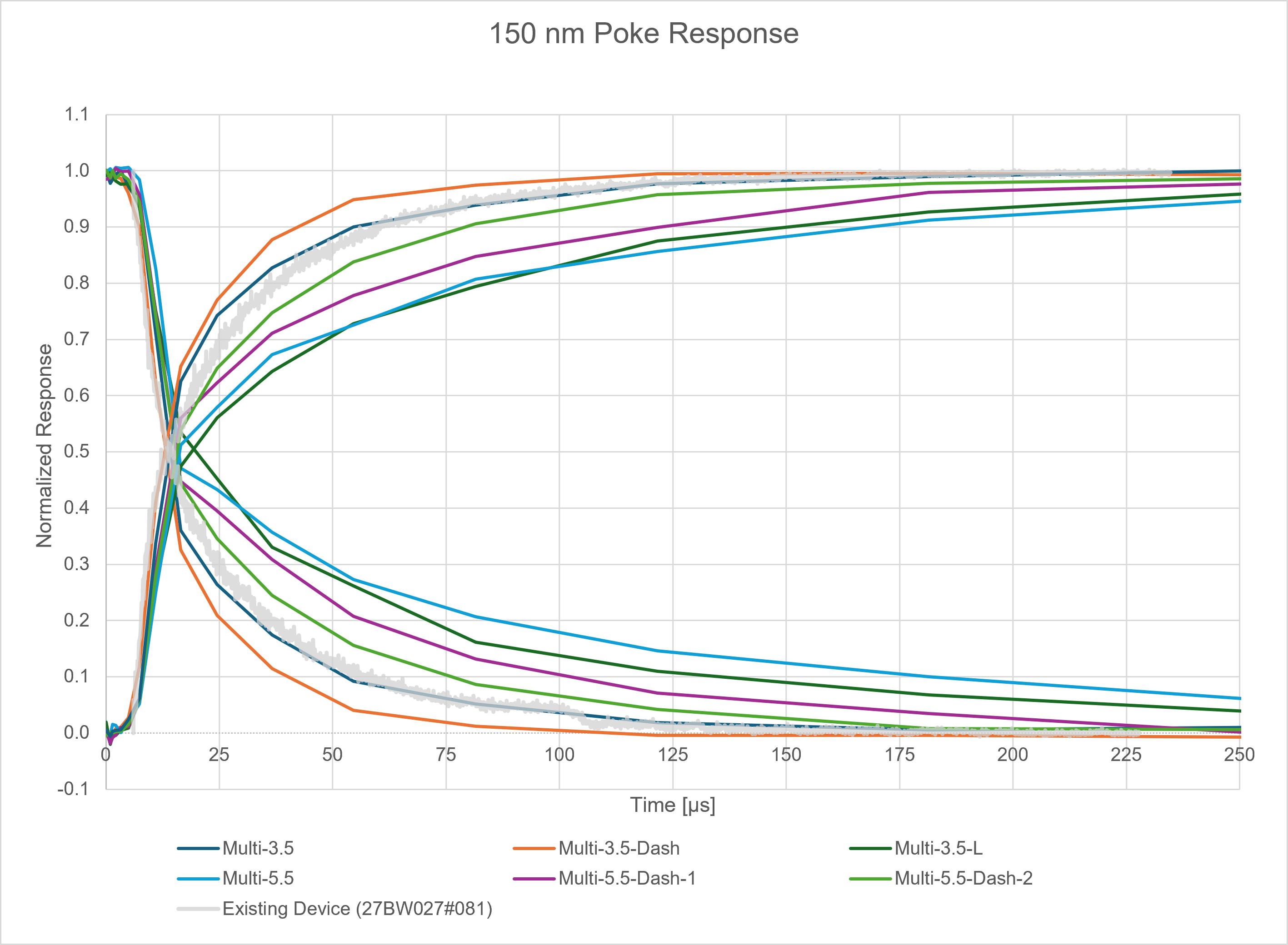}
\caption{\label{fig:dm_response} Temporal response of BMC MEMS deformable mirrors.  The light grey curve shows the response of the as-built MagAO-X device.  The new ``Multi-5.5-Dash-2'' devices (light green) will meet the $100\mu$s specification for GMagAO-X while providing higher stroke for the larger aperture of GMT.}
\end{figure}

For higher orders, we are planning to use a MEMS device with at least 5.5 $\mu$m of surface stroke.  This has necessitated a development effort in collaboration with Boston Micromachines Corporation (BMC) to improve the temporal response time of these devices.  The results are summarized in Figure \ref{fig:dm_response}, which shows that the ``Multi-5.5-Dash-2'' architecture is nearly as fast as the existing MagAO-X device and meets the GMagAO-X specification of $<100\mu$s rise/fall time.

\subsection{Wavefront Sensing and Control System}

The lower-bench contains both a Holographic Dispersed Fringe Sensor (HDFS,\cite{10.1117/1.JATIS.8.2.021513,10.1117/1.JATIS.8.2.021515}) for coarse phasing control and a 3-sided pyramid WFS (PyWFS) for higher order control.  We plan to have both visible and near-IR (through 1.8 $\mu$m) PyWFS.

Within the coronagraph, both focal-plane low-order WFS (FLOWFS) and Lyot-plane LOWFS (LLOWFS) will make use of rejected light.  To obtain high contrast we will employ focal-plane WFS techniques, such as implicit Electric Field Conjugation (iEFC)\cite{2026arXiv260708146H}.  The self-coherent camera (SCC) is also under consideration\cite{liberman_spie_2026}.

See Haffert et al.\cite{haffert_spie_2026_2} in these proceedings for a complete overview of the WFS\&C strategies.

The software system for GMagAO-X is based on the proven software suite developed for MagAO-X, now called the ``eXtreme Wavefront Control Toolkit'' (XWCTk)\cite{males_spie_2026_2}.  The XWCTk is based on the Compute and Control for Adaptive Optics (CACAO\cite{2018SPIE10703E..1EG,deo_spie_2026}) system. CACAO is in routine use at SCExAO\cite{2015PASP..127..890J} and on MagAO-X. 

The XWCTk will be adapted to interface with the GMT software and controls system, ensuring that GMagAO-X functions safely and efficiently as part of the observatory system without losing the proven heritage of MagAO-X on-sky operations.

\section{Conclusion}

GMagAO-X will be the first-light ExAO coronagraph on the GMT.  It represents the earliest opportunity of the ELT era to begin search nearby terrestrial planets for life.  GMagAO-X is now in the GMT project plan and is in the final design phase.  The scientific potential of GMagAO-X motivates a positive decision to move forward with construction of the GMT as rapidly as possible.

\acknowledgments 

The GMagAO-X conceptual and preliminary design would not have been possible without the support of the University of Arizona Space Institute.  We are also grateful for the support of an anonymous donor to Steward Observatory.  Our Final Design activities are supported by the Giant Magellan Telescope.

\bibliography{report} 
\bibliographystyle{spiebib} 

\end{document}